\documentclass{PoS}

\title{Confinement in high- temperature lattice gauge theories}

\ShortTitle{Confinement in high- temperature lattice gauge theories}

\author{\speaker{Michael Ogilvie}%
        \thanks{The author
gratefully acknowledges the support of this work by the U.S. Dept.
of Energy under Grant 91ER40628.}\\
       Washington University, St. Louis\\
       E-mail: \email{mco@physics.wustl.edu}}

\abstract{There has been substantial progress in understanding a class of SU(N)
gauge theories that are confining at high temperatures. This class
includes theories with center-symmetric Polyakov loop deformations
or with periodic adjoint fermions. The crucial role of monopoles in
lattice gauge theories of this type can be understood analytically.
The basic mechanisms occur in the two-dimensional O(3) spin model,
deformed by appropriate mass term to give an XY model. Vortices of
the XY model are constituents of O(3) instantons just as SU(N) magnetic
monopoles are constituents of KvBLL instantons. Similar methods applied
to an SU(2) lattice gauge theory yield an effective U(1) description
in which monopoles are responsible for confinement.}

\FullConference{The 30th International Symposium on Lattice Field Theory\\
		 June 24-29,  2012\\
		 Cairns, Australia}

\begin{document}





\section{High-$T$ confinement on $R^{3}\times S^{1}$}

It is now possible to construct four-dimensional gauge theories for
which confinement may be reliably demonstrated using semiclassical
methods \cite{Myers:2007vc,Unsal:2007vu}. Essentially, these models
exhibit confinement at high temperature. All of the models in this
class have one or more small compact directions, and the methods and
concepts are largely taken from finite the physics of gauge theories
at finite temperature. These models combine 
the effective potential for the Polyakov loop $P$, 
$Z(N)$ center symmetry, instantons, and
monopoles into a satisfying picture of confinement.
At temperature $T\gg\Lambda$, we have $g^2\left(T\right) \ll 1$
so semiclassical methods may
be used reliably. 
The one-loop effective potential for the Polyakov loop
shows that gauge theories are generally in the deconfined
phase at high $T$. 
However, it is possible to regain the confined phase
by modifying the action. This leads to a perturbative calculation of
possible phase structures, which turns out to be very rich, as well
as a perturbative understanding of Polyakov loop physics in the confined
phase. Furthermore, there is a
non-perturbative mechanism for confinement, as measured
by spatial Wilson loops. In this confinement
mechanism, a key role is played by finite-temperature instantons,
also known as calorons, and their monopole constituents. 

\begin{figure}
\centering\includegraphics[width=4in]{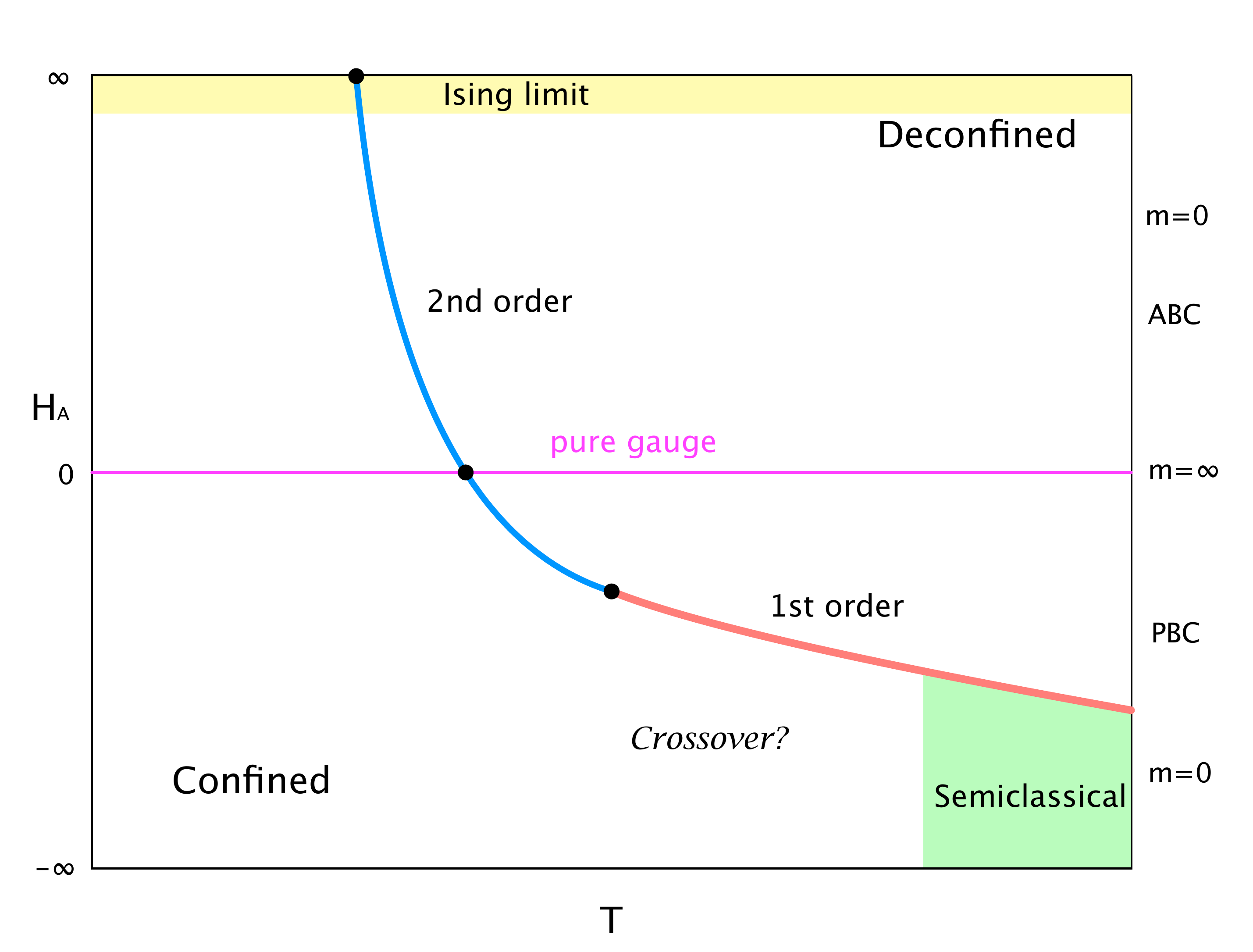}
\caption{\label{fig:su2-phase-diagram-1}$SU(2)$ phase diagram}
\end{figure}

The simplest approach to restoring confinement at
high $T$ deforms the pure gauge theory by
adding additional terms to the gauge action \cite{Myers:2007vc,Ogilvie:2007tj,Unsal:2008ch}.
For $SU(2)$, a deformation of the form 
\begin{equation}
S\rightarrow S-\beta\int d^{3}x\, H_{A}Tr_{A}P\left(\vec{x},x_{4}\right)
\end{equation}
can be used,
where the value of $x_{4}$ is arbitrary.
If the coefficient $H_{A}$ is sufficiently negative, the deformation
will counteract the effects of the one-loop effective potential, and
$Z(N)$ symmetry will hold for large $T$. 
The schematic form of the phase diagram in the $T-H_{A}$ plane for
an $SU(2)$ gauge theory with a deformation of either type is shown
in Fig. \ref{fig:su2-phase-diagram-1}. Positive values of $H_{A}$
favor $Z(2)$ symmetry-breaking, and the critical temperature will
decrease as $H_{A}$ increases. In the limit $H_{A}\rightarrow\infty$,
the Polyakov loops will only take on values in $Z(2)$; this is therefore
an Ising limit. On the other hand, negative values of $H_{A}$ favor
$Tr_{F}P=0$. This leads to a rise in the critical temperature. For
the specific deformation considered here, the critical line switches
to first-order behavior at a tricritical point. This is familiar but
non-universal behavior in $Z(2)$ models \cite{Nishimura:2011md}.
For sufficiently negative $H_{A}$, we reach the semiclassical region
where the running coupling $g(T)$ is small and semiclassical methods
may be applied reliably. The right-hand axis shows the
correspondence with an alternative approach to restoring confinement,
adjoint fermions \cite{Unsal:2007vu}, as the fermion mass
is varied.

The analysis of instanton effects in this model is based on Polyakov's study
of the Georgi-Glashow model in three dimensions \cite{Polyakov:1976fu}.
This is an $SU(2)$ gauge model coupled to an adjoint Higgs scalar,
a role that is played in the four-dimensional theory by $A_{4}$.
The monopole solutions of the field equations in four dimensions
are instantons in three dimensions.
Polyakov showed that a gas of such
three-dimensional monopoles gives rise to non-perturbative confinement
in three dimensions, even though the theory appears to be in a Higgs
phase perturbatively. This analysis carries over to
four-dimensional theories that are confined at high $T$.

\section{$O(3)$ model in $d=2$}

\begin{figure}
\centering\includegraphics[width=4in]{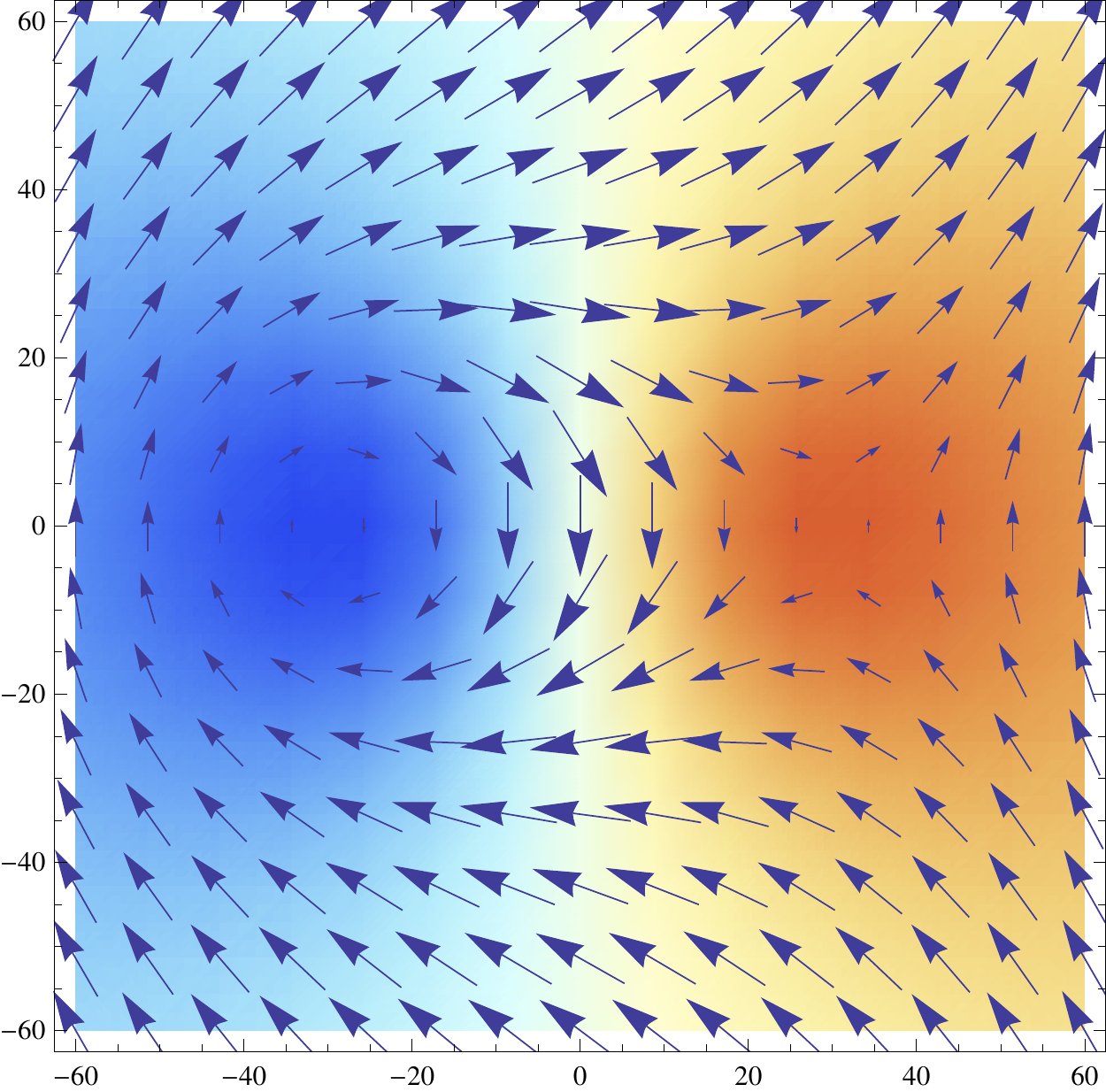}
\caption{\label{fig:Hedgehog-solution}Instanton solution in $O(3)$ model.}
\end{figure}

It is natural to ask if the continuum methods and results have parallels
in lattice gauge theory. 
We illustrate how lattice theories handle topological content
using the two-dimensional $O(3)$ model. 
 Like QCD, this is an asymptotically free theory that has instantons
\cite{Polyakov:1975rr}. As in finite temperature QCD, instantons
can be decomposed into constituents \cite{Gross:1977wu}. In the case
of the $O(3)$ model, 
these constituents can be identified with XY-model vortices \cite{Ogilvie:1981yw}.
In Fig. \ref{fig:Hedgehog-solution}, a classical instanton solution
is shown, with the arrows denoting the field components in the
$\sigma_{1}-\sigma_{2}$ plane, and the colors denoting the value
of $\sigma_{3}$. The embedding of the vortex-antivortex solution
within the instanton is obvious.
The $O(3)$ model can be deformed into an XY model by the addition
of a mass term for $\sigma_{3}$ \cite{Ogilvie:1981yw,Satija:1982ja,Affleck:1985jy}:
\begin{equation}
S\rightarrow S+\int d^{2}x\,\frac{1}{2}h\sigma_{3}^{2}
\end{equation}
in a manner similar to finite-temperature QCD. The mass term breaks
the classical conformance invariance of the model and makes it effectively
Abelian at large distances. It is physically obvious that as $h$
increases, the deformed $O(3)$ model will become more and more like
an XY model, and the constituent vortices inside instantons should
be identified with the Kosterlitz-Thouless vortices of the XY model.
To make this identification precise, we consider a lattice form of
the deformed $O(3)$ model.

The lattice action is given by
\begin{equation}
S=-\sum_{x,\mu}K\vec\sigma(x)\cdot\vec\sigma(x+\mu)+\sum_{x}\frac{1}{2}h\sigma_{3}^{2}(x)
\end{equation}
 where $x$ is now a lattice site and $\mu$ one of two lattice directions.
We parametrize $\vec{\sigma}$ as
\begin{equation}
\vec{\sigma}=\left(\sqrt{1-\sigma_{3}^{2}}\cos\theta,\sqrt{1-\sigma_{3}^{2}}\sin\theta,\sigma_{3}\right).
\end{equation}
We can decompose the action as
\begin{equation}
S=-\sum_{x,\mu}K_{eff}\left(x,\mu\right)\cos\left[\theta\left(x\right)-\theta\left(x+\mu\right)\right]+S_{3}
\end{equation}
where
\begin{equation}
K_{eff}\left(x,\mu\right)= K\sqrt{1-\sigma_{3}^{2}\left(x\right)}\sqrt{1-\sigma_{3}^{2}\left(x+\mu\right)}
\end{equation}
and
\begin{equation}
S_{3}=-\sum_{x,\mu}K\sigma_{3}(x)\sigma_{3}(x+\mu)+\sum_{x}\frac{1}{2}h\sigma_{3}^{2}(x)
\end{equation}
both depend only on $\sigma_{3}$.

At this point, we can follow well-known arguments
to obtain a form for the model that explicitly includes vortex
effects \cite{Jose:1977gm}. 
We write the partition function as
\begin{equation}
Z=\int_{S^{2}}\left[d\sigma\right]e^{-S}=\int_{-1}^{+1}\left[d\sigma_{3}\left(x\right)\right]e^{-S_{3}}\int_{S^{1}}\left[d\theta\right]\prod_{x,\mu}e^{K_{eff}\left(x,\mu\right)\cos\left(\nabla_{\mu}\theta\left(x\right)\right)}
\end{equation}
where $\nabla_{\mu}\theta\left(x\right)\equiv\theta(x+\mu)-\theta(x)$.
For each link, we expand the interaction in a character expansion,
which is a Fourier series: 
\begin{equation}
Z=\int_{-1}^{+1}\left[d\sigma_{3}\left(x\right)\right]e^{S_{3}}\int_{S^{1}}\left[d\theta\right]\prod_{x,\mu}\sum_{n_{\mu}\left(x\right)\in Z}I_{n_{\mu}\left(x\right)}(K_{eff}\left(x,\mu\right))e^{in_{\mu}\left(x\right)\nabla_{\mu}\theta\left(x\right)}
\end{equation}
where $I_{n}$ is a modified Bessel function. This step introduces
integer variables $n_{\mu}\left(x\right)$ on every link. We now make
use of the asymptotic form of $I_{n}$ for $K_{eff}\gg1$, using what
is called the Villain approximation, obtaining
\begin{equation}
Z=\int_{-1}^{+1}\left[d\sigma_{3}\left(x\right)\right]e^{S_{3}}\int_{S^{1}}\left[d\theta\right]\prod_{x,\mu}\sum_{n_{\mu}\left(x\right)\in Z}\frac{1}{\sqrt{2\pi K_{eff}\left(x,\mu\right)}}e^{K_{eff}\left(x,\mu\right)-n_{\mu}^{2}\left(x\right)/2K_{eff}\left(x,\mu\right)}e^{in_{\mu}\left(x\right)\nabla_{\mu}\theta\left(x\right)}
\end{equation}
Although this step appears here as an approximation, it is really
a small deformation of the action that does not change the critical
properties of the model. It is now easy to integrate over the $\theta$
variables, which leads to the constraint$ $$\nabla_{\mu}n_{\mu}\left(x\right)=0$.
This in turns allows us to write $n_{\mu}(x)=\epsilon_{\mu\nu}\nabla_{\nu}m(X)$
where $m\left(X\right)$ is an integer-valued field on the dual lattice
site $X$ which is displaced from $x$ by half a lattice spacing in
each direction. The partition function is now
\begin{equation}
Z=\int_{-1}^{+1}\left[d\sigma_{3}\left(x\right)\right]e^{-S'_{3}}\sum_{\{m\left(X\right)\}\in Z}e^{-\sum_{X,\nu}\left(\nabla_{\nu}m\left(X\right)\right)^{2}/2K_{eff}\left(x,\mu\right)}
\end{equation}
where
\begin{equation}
S'_{3}=S_{3}-\sum_{x,\mu}\left[K_{eff}\left(x,\mu\right)-\frac{1}{2}\log\left(2\pi K_{eff}\left(x,\mu\right)\right)\right]
\end{equation}
The final step is to introduce a new field $\phi(x)\in R$ using a
periodic $\delta$-function, effectively performing a Poisson resummation:
\begin{equation}
Z=\int_{-1}^{+1}\left[d\sigma_{3}\left(x\right)\right]e^{S'_{3}}\int_{R}\left[d\phi\left(X\right)\right]e^{-\sum_{X,\nu}\left(\nabla_{\nu}\phi\left(X\right)\right)^{2}/2K_{eff}\left(x,\mu\right)}\sum_{\{m\left(X\right)\}\in Z}e^{2\pi im\left(X\right)\phi\left(X\right)}.
\end{equation}

We see from this form of the partition function that vortices are
explicitly present in the functional integral, induced by source $m(X)$
on the dual lattice. For each configuration $\{m\left(X\right)\}$,
the integral over $\phi$ and $\sigma_{3}$ must be carried out. This
can be done using standard perturbative methods. Each dual lattice
site $X$ where $m(X)\ne0$, will be the site of a vortex of charge
$m(X)$. In a dilute gas approximation, we can see that the size of
the vortex core will in general be set by the scale-setting parameter
$h$, which determines the region around $X$ where $\sigma_{3}$
is significantly different from zero. The contribution of the vortex
core to the total weight of a given configuration $\{m\left(X\right)\}$
can be captured in a vortex activity $y$, which represents the Boltzmann
weight of the classical lattice vortex solution times a functional determinant
factor, just as in the continuum. It is clear that in the limit where
$h$ is very large, $\sigma_{3}$ will be essentially zero everywhere,
and we recover the $XY$ model with $K_{eff}\simeq K$ and a vortex
core size on the order of the lattice spacing. Note that the $Z(2)$
symmetry under $\sigma_{3}\rightarrow-\sigma_{3}$ means that for
each vortex winding number $m$, there are two types of vortices depending
on the behavior of $\sigma_{3}$ in the core, as in the continuum.
For $h>0$, the large-distance behavior is that of an XY model, giving
a continuous path between the $O(3)$ model and the vortex Coulomb
gas phase of the XY model. If we keep only the $m=1$ contributions,
we have essentially a lattice sine-Gordon model
\begin{equation}
Z=\int_{R}\left[d\phi\left(X\right)\right]\exp\left[-\sum_{X,\mu}\frac{1}{2\bar{K}_{eff}}\left(\nabla_{\mu}\phi\left(X\right)\right)^{2}+\sum_{X}4y\cos\left(2\pi\phi\left(X\right)\right)\right]
\end{equation}
where $\bar{K}_{eff}$ is the value of $K_{eff}$ away from the vortex
cores. All of the physics associated with the short-ranged $\sigma_{3}$
field is contained in $\bar{K}_{eff}$ and $y$.

\section{High-T confinement for lattice $SU(2)$ in $d=4$}

The $(3+1)$-dimensional $SU(2)$ gauge theory at high temperatures
can be treated in much the same way as the two-dimensional $O(3)$
model. It is convenient to work in Polyakov gauge, where $A_{4}$
is diagonal and time-independent so that the Polyakov loop is given
by $P=\exp\left(iA_{4}/T\right)=\exp\left(i\theta\tau_{3}\right)$.
A sufficiently strong deformation term will make the expected
value of the timelike link variable $U_{4}=\exp(iA_{4})$ significantly
different from one. This in term will give large masses to the off-diagonal
parts of the $U_{j}$ fields. The off-diagonal fields will be important
only inside monopole cores where $A_{4}$ is small. Outside monopole
cores, the model is effectively Abelian. 

A simplified approach is to take the deformation term to be very strong
and assume that all the fields are independent of $x_{4}$. We take
the the timelike links $U_{4}\left(\vec{x,}t\right)$ to be diagonal
and independent of $t$: $U_{4}\left(\vec{x}\right)=\exp\left(i\tau_3\theta_{0}\left(\vec{x}\right)\right)$.
A strong deformation term forces $\left\langle Tr_{F}P\right\rangle =0$
with an expected value for $\left\langle \theta_{0}\right\rangle $,
given by $N_{t}\left\langle \theta_{0}\right\rangle =\pi/2$. As in
the $O(3)$ case, we can define the (dimensionally-reduced) spatial
gauge fields as 
\begin{equation}
U_{j}\left(\vec{x}\right)=\sqrt{1-\left(U_{j}^{1}\left(\vec{x}\right)\right)^{2}-\left(U_{j}^{2}\left(\vec{x}\right)\right)^{2}}
\exp\left(i\tau_3\theta_{j}\left(\vec{x}\right)\right)
+i\tau_{1}\cdot U_{j}^{1}\left(\vec{x}\right)+i\tau_{2}\cdot U_{j}^{2}\left(\vec{x}\right)
\end{equation}
The expectation value $\left\langle U_{0}\right\rangle $ makes the
$U_{j}^{1}$ and $U_{j}^{2}$ fields massive, and they do not contribute
to the large-distance behavior. This leaves us with an effective three-dimensional
$U(1)$ gauge theory. The dual of a a three-dimensional Abelian gauge
theory is an Abelian spin system, in this case again yielding a lattice
sine-Gordon model as in the continuum \cite{Banks:1977cc}.

The above simplified approach, based on the early application of dimensional
reduction, is in fact too simple. As in the $O(3)$ model, where there
were two types of vortices and two types of antivortices distinguished
by their behavior in the vortex core, there are four Eucldean monopole
solutions, not two \cite{Kraan:1998kp,Kraan:1998pm,Lee:1998bb,Davies:1999uw}.
The BPS-type monopole and anti-monopole solutions can be constructed
as conventional time-independent monopole solutions, and are thus
included in the simplified approach. On the other hand, the KK-type
solutions are constructed from the BPS solutions using an $x_{4}$-dependent,
non-periodic gauge transformation that changes the instanton charge
of a field configuration\cite{Davies:1999uw}. Thus, a proper treatment
of both types of monopoles is necessary. After accounting carefully
for both types of solutions, the dual form of the partition function
in the confined phase is
\begin{equation}
Z=\int_{R}\left[d\phi\left(X\right)\right]\exp\left[-\sum_{X,\mu}\frac{g^{2}}{8N_{t}}\left(\nabla_{\mu}\phi\left(X\right)\right)^{2}+\sum_{X}4y\cos\left(2\pi\phi\left(X\right)\right)\right]
\end{equation}
which has the same form as the corresponding continuum result, where
the effective action has the form
\begin{equation}
S_{eff}=\int d^{3}x\left[\frac{g^{2}(T)T}{32\pi^{2}}\left(\partial_{j}\phi\right)^{2}-4y\cos(\phi)\right].
\end{equation}
The two results are equivalent after identifying $T^{-1}$ with $N_{t}$
and rescaling the $\phi$ field.

\section{Conclusions}

We have seen that non-Abelian lattice theories deformed to Abelian
effective theories work in the same way as their continuum counterparts,
and lattice duality reproduces semiclassical continuum duality in
these cases. In the case of $SU(2)$ lattice gauge theory, there is
a continuous path between the confined phase of SU(2) and the monopole-dominated
phase of lattice U(1) gauge theory. The lattice theories know something
about continuum topology, but care must be exercised. One particularly
interesting feature of the lattice analysis is that the construction
works for $O(N)$ models with $N>3$ when a term of the form
\begin{equation}
\sum_{x}\frac{1}{2}h\sum_{j=3}^{N}\sigma_{j}^{2}(x)
\end{equation}
is added to the lattice action. It is clear on physical grounds that
all the $O(N)$ models should reduce to an XY model at large distances
when deformed appropriately. However, instantons only appear in the
$O(3)$ model. It is apparent that not all non-perturbative contributions
are clearly associated with instantons and that instanton methods
are in some sense incomplete prescriptions for determining the non-perturbative
content of a theory; see the recent work of Argyres and Unsal for
a related continuum perspective \cite{Argyres:2012vv,Argyres:2012ka}.


\end{document}